# Next generation combined sonic-hotfilm anemometer: wind alignment and automated calibration procedure using deep learning

Roni H. Goldshmid[1,2], Ewelina Winiarska[1], and Dan Liberzon[1]

**Abstract**

The study of naturally occurring turbulent flows requires ability to collect empirical data down to the fine scales. While hotwire anemometry offers such ability, the open field studies are uncommon due to the cumbersome calibration procedure and operational requirements of hotwire anemometry, e.g., constant ambient properties and steady flow conditions. The combo probe—the combined sonic-hotfilm anemometer—developed and tested over the last decade has demonstrated its ability to overcome this hurdle. The older generation had a limited wind alignment range of 120 degrees and the in-situ calibration procedure was human decision based. This study presents the next generation of the combo probe design, and the new fully automated in-situ calibration procedure implementing deep learning. The elegant new design now enables measurements of the incoming wind flow in a 360-degree range. The improved calibration procedure is shown to have the robustness necessary for operation in everchanging open field flow and environmental conditions. This is especially useful with diurnally changing environments or non-stationary measuring stations, i.e., probes placed on moving platforms like boats, drones, and weather balloons. Together, the updated design and the new calibration procedure, allow for continuous field measurements with minimal to no human interaction, enabling near real-time monitoring of fine-scale turbulent fluctuations. Integration of these probes will contribute toward generation of a large pool of field data to be collected to unravel the intricacies of all scales of turbulent flows occurring in natural setups.

## 1 Introduction

An ability to obtain empirical field records of large to fine scale velocity field fluctuations is crucial for improving the physical understanding of the intricacies of turbulent flows. However, fine scale field measurements of turbulent fluctuations are less common due to low spatiotemporal resolution of widely used anemometers, such as ultrasonic anemometers (sonics) and LiDARs. The use of hotwire (or hotfilms) anemometry in the field is limited mainly due to the wind direction and temperature variations, forcing frequent and complex re-calibration (Skelly et al. 2002; Nelson et al. 2011). The collocated sonic-hotfilm anemometer, dubbed the combo probe, was designed to overcome this hurdle, Figure 1. Developed and implemented in several field setups during the last decade (Kit et al. 2010, 2017; Kit and Grits 2011; Vitkin et al. 2014; Kit and Liberzon 2016; Goldshmid and Liberzon 2018), the combo probe is capable of continuously sensing the fine-scales of turbulent fluctuations in the field, without the need for human interaction for frequent cumbersome re-calibrations. Consisting of a collocated sonic and two x-shaped double sensor hotfilm probes, the combo simultaneously samples the hotfilms and sonic at a high sampling frequency of several kilohertz. Machine Learning is used post-measurements to produce accurate calibration for the hotfilm voltages by implementing artificial Neural Networks (NN): the voltages are calibrated against low resolution sonic anemometer data. More specifically, the combo successfully overcomes two major aspects that pose a challenge when using hotwires in the field, but still had several limitations that we tackle and omit here.

The *first aspect* of hotwire anemometry that posed a challenge in field measurements is the need for a repetitive cumbersome calibration. The calibration of the wires typically involves a low turbulence flow (e.g., a controlled jet or a wind tunnel) and preferably a pitch/yaw manipulator to calibrate for a wide range of angles of attack. Once completed, it is only valid for a narrow range of varying environmental conditions of temperature and humidity. When a significant enough change occurs, a recalibration is required. Since the change of environmental conditions in the field is inevitable, the combo probe tackles this limitation utilizing an alternative approach to the traditional hotwire anemometry calibration procedure. It uses the pre-calibrated sonic data to train NN as an the in-situ calibration function. This in-situ calibration procedure was developed by Kit et. al. (Kit et al. 2010) and tested in several studies (Fernando et al. 2015; Kit et al. 2017; Goldshmid and Liberzon 2018, 2020) proving its robustness for use in various environmental conditions. Usually divided into hourly sets of close-to-steady conditions, it involves a careful human-decision based selection of five representative minutes from

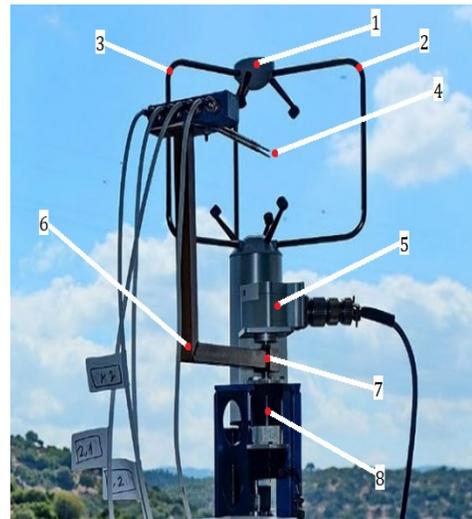

Figure 1 Previous configuration (Kit et al. 2010) of the combo probe. This image is taken during the field experiment in Nofit, Israel (Goldshmid and Liberzon 2018). The main components of the combo probe are 1: ultrasonic anemometer. 2-3: sonic support struts limiting the hotfilm sensors orientation within 120° range. 4: hotfilm sensors. 5: absolute encoder. 6: rotating arm to align the sensors. 7: center of rotation of the arm. 8: motor and gearhead.

---

each hour of continuous measurements to form the training set (TS) that would be valid for all data collected within that hour. The selected minutes are to represent the entire range of mean velocity variations of the hour and have high quadratic mean (rms) values to also represent the instantaneous velocity fluctuations as best as possible. Two separate NN are trained for every hour, one for each x-shaped hotfilm sensor, which are used as the voltage to velocity transfer functions. After the TS minutes are selected, the corresponding sonic velocities and HF voltages data undergo low pass filtering down to the trusted frequency of the sonic. The filtered velocity data from the sonic are used as targets while the filtered voltage data from the hotfilms are used as inputs. Finally, the original high-frequency HF voltages are fed into the trained NN, producing records of 3D velocity fluctuations field at high spatiotemporal resolution. Automation of this procedure, specifically omitting the need for manual selection of the TS minutes, would allow the combo to operate remotely, and this study proposes an approach to achieve this.

The *second aspect* of hotwire anemometry that poses a challenge for field measurements is the requirement that the probe remains aligned with the mean flow direction, as the accuracy of the hotwire anemometry is proportional to the probe alignment with the mean flow (Bruun 1995). Unlike in a controlled laboratory setup, the wind direction in the field is everchanging, requiring frequent realignment of the sensors to maintain the mean flow angle of attack within predefined tight boundaries. The combo tackles this limitation by mounting the hotfilm sensors on a rotating arm and using a software routine for periodic realignment of the sensors with the mean flow direction. The sonic provided velocity field at the end of every measurement (usually minute long) interval determines the new desired probe direction, while a motor and encoder are used to rotate the arm. In the post-processing, outlier minutes with a significantly varying flow direction or angles of attack larger than $\pm 10°$, are omitted. The combo is limited to measuring range of 120° due to the sonic support struts obstacle and the new combo design tackles this limitation.

This paper presents the two advancements made to the combo anemometer: a new mechanical design and an automated calibration procedure. It is organized as follows: the new mechanical design is described in Section 2.1. It includes a more compact design of the holding and rotating mechanism to allow accurate constant realignment of the hotfilm probe with the mean flow. The new rotating mechanism can complete 360° turns, while assuring the hotfilm probe is always positioned at the rotation center collocated with the sonic control volume. The testing of the new combo design was conducted in both a controllable and noncontrollable environment, detailed in Sections 2.2 and 2.3 respectively. More specifically, Section 2.2 describes the laboratory experimental setup - a large environmental wind tunnel in which a turbulent boundary layer (BL) is created. While Section 2.3 describes the field experimental setup – wind measurements in marine atmospheric boundary layer (MABL) at the Gulf of Aqaba (Red Sea). Section 2.4 details the changes proposed to the NN based calibration procedure, aimed to achieve a complete automation of the calibration. The automated procedure enables (almost) real-time data processing, which would allow the use of combo probes in meteorological stations with the ability to monitor real-time fine scale turbulence statistics and input them into prediction models. Finally, Section 3 presents the results obtained in the wind tunnel using the automated in-situ calibration method and the new calibration methodology testing on field data; it also presents an in-depth discussion of the findings. Altogether, the new mechanical design and new automated calibration procedure ensure persistent accuracy in the absence of human decision factor, ensure robust operation over prolonged periods, enable close-to-real time data processing making it suitable for meteorological stations and on non-stationary platforms such as moving terrain or marine vehicle, planes, and UAVs. While previous works demonstrated the combo ability to provide accurate measurements of the fine turbulent flow scales in both stable and unstable BL flows (Fernando et al. 2015; Kit et al. 2017; Goldshmid and Liberzon 2018, 2020), this paper provides a brief preview of evidence that the new combo design can even be deployed in the open field/sea and measure the wind coming from all directions in unsteady environments.

## 2 Methods

### 2.1 New mechanical design

Building on the previous configuration of the combo anemometer, demonstrated in Figure 1, this section details the most recent mechanical design advancements accomplished for improving the combo usability in field measurements. The previous design of the combo had two mechanical constraints. The first is that the hotfilm probe orientation range was restricted by the sonic support struts, labels 2 and 3 from Figure 1, limiting the realignment of the probe within a 120° range. The second is the location of the axis of rotation of the hotfilm probe (the two x-shaped sensors). The positioning of the motor and the gearhead is several centimeters from the center of the sonic control volume, causing the hotfilm probe to be at different distances from the center of the sonic measurement volume at each yaw angle. While the latter limitation was shown not to cause a major reduction in accuracy, the former limitation did pose a significant constrain on the ability to monitor and record wind flow during diurnal and weekly changes in mean wind direction. Both limitations are fully addressed in the here presented new design.

The previous combo configuration had the sonic fixed in place facing north (as per the convention) and the hotfilm probe rotating in the sonic control volume to face the mean direction of the flow. When the data was collected, it was all converted to correspond with the coordinate system of the hotfilm probe. The newly proposed configuration suggests rigidly mounting the hotfilm probe to the sonic and rotating them together according to the direction of the flow. The mean wind direction is determined by the sonic provided velocity field components, and the motor realigns the combo. Keeping track of the sonic and probe orientation relative to north is achievable using an absolute encoder.

Unlike the older design of the combo, the new one enables coverage of the complete 360° range of wind. The entire combo is rotated on a Schneider Electric Motor with a built-in absolute encoder, model LMDAE853. A more detailed image of the mechanics is presented in Figure 2. It details the elegant, more rigid, and eventually simpler construction and operational combo design. Eliminating the need for coordinate systems translation the new design also simplifies the control software routine, therefore minimizing the chance of error or damage to the hotfilms.

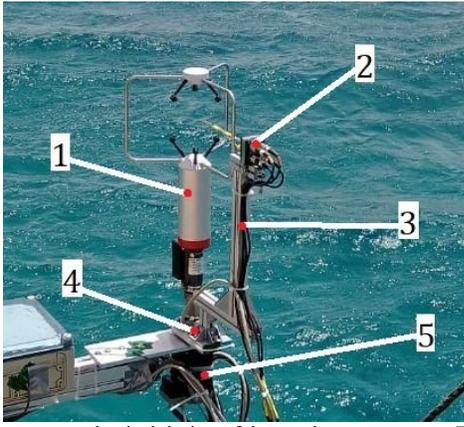

Figure 2 The new mechanical design of the combo anemometer. The combo, sonic and hotfilm probes, rotates as a single unit according to the mean direction of the flow determined by the sonic. This newest version of the combo is composed of the following 1: sonic. 2: two x-shaped hotfilm sensors (hotfilm probe). 3: aluminum profile probe holder (arm). 4: bottom mounting pate to supporting the sonic and the hotfilm holder/arm. 5: motor with an absolute encoder.

## 2.2 Wind tunnel experimental setup

The laboratory experimental setup took place in the Environmental Wind Tunnel at the Civil and Environmental Engineering Faculty of the Technion-Israel Institute of Technology. This wind tunnel (Figure 3) has a $12\ m$ long test section that operates using a powerful blower in an open-circuit flow-suction mode that can produce mean wind velocities of up to $14\ m/s$. The cross section is a square of $2.0\ m$ on each side and the roof is adjustable to enable control of pressure gradients in the streamwise direction. The wind tunnel inlet is equipped with a flow straightening honeycomb that is $15\ cm$ long and consists of cylinders with a $5\ cm$ diameter.

The wind tunnel currently hosts a physical scaled model of a corn field used for a project that examines wind interactions with vegetation canopy. We used the existing setup to compare the hotwire calibration procedures by deploying the combo in the wind tunnel BL; the hotwires position was $55\ cm$ from the wind tunnel floor (see Figure 3c) and their position along the wind tunnel main axis is depicted by the red circled x in Figure 3b. The corn field model and the airflow BL generation elements were constructed based on literature reviews and some trial and error. The design we used included a grid with various densities to generate shear flow and artificially increase the BL height—it is most dense at the bottom and least dense at the top. The grid is positioned between the vortex generating spires and the model canopy, while gravel is spread in the entire area leading to the model canopy to provide surface roughness.

Prior to actual measurements the mean velocity profiles at the combo location for different flow rates provided by the blower were measured using pitot tubes, providing BL thickness and to confirm correct placement of the hotfilm inside the BL. The profiles for high and low flow rates of all examined experimental setups are presented in Figure 4. The velocity profiles make it evident that the hotfilm probe was indeed inside the BL for all experimental regimes. The length of the sonic acoustic fly path is $15\ cm$, and the BL height is about $70\ cm$. The ratio of the sonic length to the height of the
BL is not negligible in this case, but this is not the case in the field. This is important because the sonic averages velocities over different regions in this BL inherently introducing errors in the sonic readings. The three types of experimental setups are considered: (1) the complete setup described in Figure 3b and referred to as YGYC, which stands for yes-grid-yes-canopy; (2) the same setup as before but without the canopy model and referred to as YGNC, which stands for yes-grid-no-canopy; finally, (3) the same as before but excluding the grid, i.e. only the gravel, spires, and

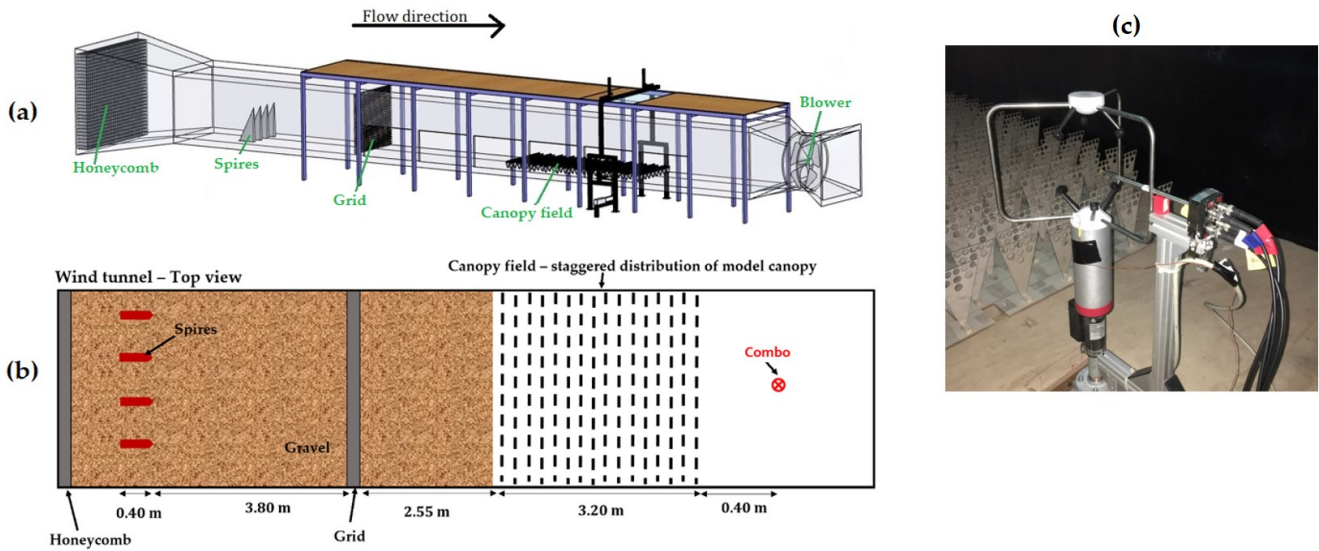

Figure 3 (a) A sketch of the Environmental Wind Tunnel at the Civil and Environmental Engineering Faculty of the Technion-Israel Institute of Technology. A suction type wind tunnel capable of up to $14\ m/s$ mean velocity. (b) The experimental setup of the artificially designed BL in the wind tunnel includes the honeycomb at the inlet on the left. Gravel covers the tunnel floor between the honeycomb and the model canopy. On the gravel are the vortex generating spires and the shear generating grid. (c) Combo in the wind tunnel is installed downstream from the model canopy.

honeycomb and referred to as NGNC, which stands for no-grid-no-canopy. The three setups result in different shape of the BL with various levels of turbulence intensity at the location of the measurement. Table 1 summarizes the three configurations examined.

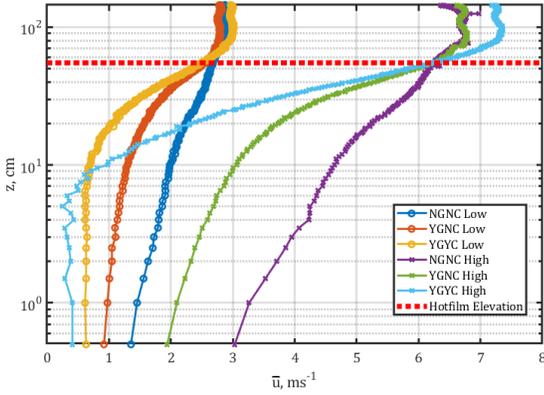

Figure 4 Profiles of the mean velocity at high and low flowrates for all three configurations examined using the combo.

Table 1 Experimental configurations and naming of wind tunnel tests

| Experiment name | Grid installed | Canopy | Honeycomb, spires, and gravel |
|---|---|---|---|
| NGNC | No | No | Yes |
| YGNC | Yes | No | Yes |
| YGYC | Yes | Yes | Yes |

After examining the mean properties of the experimental configurations, the fine scale turbulent properties are considered. The fine scales were captured using the 2 x-shaped hotfilm sensors model 1241-20W. Four miniCTA channels fabricated by Dantec Dynamics model 54T42 were used with an overheat ration of 1.7. The ultrasonic anemometer used in the combo is fabricated by RM Young, model 81000. It is common practice for controlled laboratory experiments with hotfilms to calibrate the hotwires using a low turbulence jet and a mechanical manipulator. The hotfilm calibration was performed using an automated manipulator fabricated by Dantec Dynamics and consisting of a controlled jet with a motorized pitch and yaw manipulator. The data collection and processing were performed using a specially written MATLAB routine to achieve a wider range of angles of attack and velocities for the four wires simultaneously. The jet-based calibration was performed twice, before and after the measurements took place inside the wind tunnel. Expecting high instantaneous attack angles due to high turbulence intensity in the BL, the calibration was in the shape of a wide cone covering changes in azimuth and elevation in the range of $-45° \div +45°$ with increments of $7.5°$ on each rotation axis. Covered range of mean velocities was $0 \div 15\ m/s$, each point was sampled at $6000\ Hz$ and averaged over 5 seconds. This calibration was performed twice, before and after the measurements in the wind tunnel, to correct for possible drift errors. The transfer function estimates were obtained using the lookup table method, as it was suggested to be the most accurate up-to-date (Van Dijk and Nieuwstadt 2004).

In the wind tunnel, the combo obtained data was saved in chunks of 52 seconds, like it would be done in the field operation to allow 8 seconds for realignment of the probe. The background temperature range of the calibration and flow was between $20 - 26°C$.

The spectral shapes, turbulence intensities, and various length scales of the measured turbulent BL flow are presented in Table 2. The turbulence intensity is defined as,

$$TI = \frac{\sqrt{TKE}}{\bar{u}}, \quad (1)$$

where $\bar{u}$ is the mean streamwise velocity field component, and the turbulent kinetic energy is defined as,

$$TKE = \frac{(u^2 + v^2 + w^2)}{2}, \quad (2)$$

where the $u, v, w$ represent the instantaneous fluctuations of the 3D velocity field components. The horizontal length scale is defined as,

$$L_H = \frac{\left(\sqrt{\overline{u^2}}\right)^3}{\bar{\varepsilon}}, \quad (3)$$

here $\bar{\varepsilon}$ is the TKE dissipation rate,

$$\bar{\varepsilon} = \frac{\varepsilon_u + \varepsilon_v + \varepsilon_w}{3}, \quad (4)$$

where,

$$\varepsilon_u = 15 \frac{\nu}{\bar{u}^2} \overline{\left(\frac{\partial u}{\partial t}\right)^2}, \quad (5)$$

$$\varepsilon_v = 7.5 \frac{\nu}{\bar{u}^2} \overline{\left(\frac{\partial v}{\partial t}\right)^2}, \quad (6)$$

$$\varepsilon_w = 7.5 \frac{\nu}{\bar{u}^2} \overline{\left(\frac{\partial w}{\partial t}\right)^2}. \quad (7)$$

The Taylor length scale is defined as,

$$\lambda = \sqrt{\frac{\overline{u^2}}{\overline{\left(-\frac{1}{\bar{u}}\frac{\partial u}{\partial t}\right)^2}}}, \quad (8)$$

and finally, the Kolmogorov length scale is,

$$\eta = (\nu^3/\bar{\varepsilon})^{1/4}. \quad (9)$$

The spectral shapes of the velocity field components fluctuations and length scales presented in Figure 5 assure that the BL behavior was well captured, the probe size selection was sufficient, and that the measurements were conducted properly. The spectra clearly show the three expected energy cascade ranges: the energy containing range, the inertial subrange (with close to $-5/3$ slope), and the dissipation range. Flattening out of the spectral shapes at the highest frequencies indicated that the signal to noise is too low above a specific frequency. The flattening is observed at different frequencies, ranging from $500\ Hz - 3000\ Hz$, depending on the flow regime and mean velocity.

The overheat ratio of 1.7 was sufficient but could be increased depending on the flow regime. These results further emphasize that the compensation of the spatial resolution due to the use of two independent x-probes together did not affect the measurements. As the flow becomes more turbulent the dissipation range extends to higher frequencies, manifesting the existence of smaller scales in the flow not dissipated by viscosity. The black $-5/3$ line is drawn in the same place in both subplots of Figure 5, indicating the fluctuations are also intensified with the increased experimental setup complexity.

Both flow regimes without the model canopy appear to be characterized by similar length scales values, and when the canopy is introduced the length scales appear smaller. As expected, the Taylor length scales appear to correspond scales within the inertial subrange. The Kolmogorov length scales are the largest scales in which the viscous dissipation dominates the turbulent kinetic energy and converts the energy into heat, and their values indicate that an increase in the sampling frequency and overheat ratio might provide additional information.

### 2.3 Gulf of Aqaba experimental setup

The new combo was also deployed for field measurements with the goals of testing its durability and its capability of continuous operation without human intervention for days or weeks at a time. The testing took place in an open sea environment at the northern tip of the Gulf of Aqaba (Red Sea) in May of 2019. The gulf hosts a plethora of recreational activities due to the presence a coral reef. It is also a home of several large ports since Egypt, Israel, and Jordan share its shoreline. Despite significant amount of wave activity and appealing topography that generates a diurnally repetitive wind flow along the golf, its wind-wave regime was never thoroughly investigated. The research goal of this 2019 Study was to understand the fine scale turbulent fluctuations in the local MABL. This complete 2019 Study was a more comprehensive one than the previous 2017 Study (Shani-Zerbib et al. 2019), because it included the air flow turbulence measurements using the new combo on top of the experimental setup used also in 2017 (Shani-Zerbib et al. 2018), (Shani-Zerbib and Liberzon 2018). Only a small portion of the results of the 2019 Study will be presented here; it is only for the sake of comparison of the newly proposed automated calibration procedure with the previously used manual calibration procedure.

The measurements took place at a pier designed for marine monitoring at the Interuniversity Institute for Marine Sciences (IUI) in Eilat, Israel, situated in the southmost part of Israel. Figure 6 presents the unique topography of the gulf that makes it an appealing location for wind-wave interaction research. It is located between relatively high mountain ridges that form an almost rectangular shape of 15 $km$ width and is composed of relatively deep water. The rectangular shape of the gulf has a 6.5 $km$ fetch of deep water up to the point of measurements. It also hosts a rather unique and steady diurnal wind pattern during daylight hours, characterized as approximately $10° - 20°$ northeastern wind of $8 - 12\ m/s$.

Table 2 Turbulence statistics of the examined flow in the wind tunnel.

| Flow rate | Experimental configuration | $TI$, % | $L_H$, $\times 10^{-1} m$ | $\lambda$, $\times 10^{-3} m$ | $\eta$, $\times 10^{-4} m$ |
|---|---|---|---|---|---|
| Low | NGNC | 6.0 | 0.2 | 5.59 | 3.4 |
|  | YGNC | 13 | 1.7 | 10.7 | 3.4 |
|  | YGYC | 31 | 0.5 | 4.38 | 1.9 |
| High | NGNC | 6.0 | 4.7 | 13.4 | 3.7 |
|  | YGNC | 10 | 5.0 | 13.5 | 2.9 |
|  | YGYC | 22 | 0.9 | 7.40 | 1.4 |

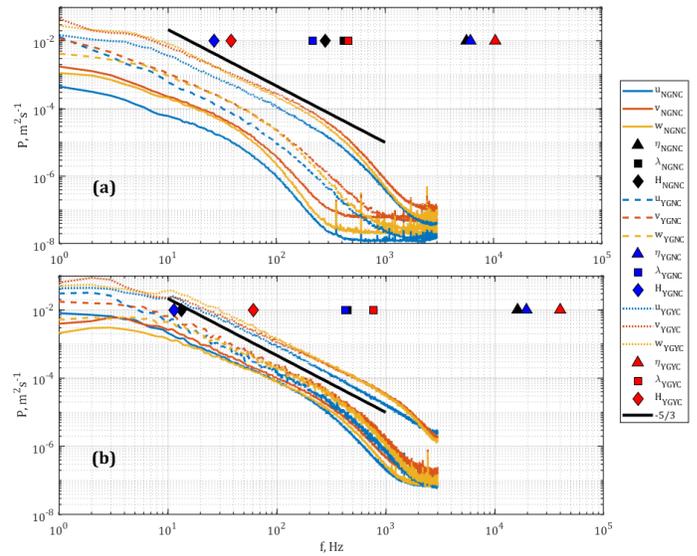

Figure 5 Power density spectra of velocity components fluctuations, with the *low* and *high* flow rates examined. The results presented here provide the three experimental setups of YGYC, YGNC, and NGNC. The horizontal length scale, Taylor length scale, ad Kolmogorov length scale for each of the experimental configuration are presented as well.

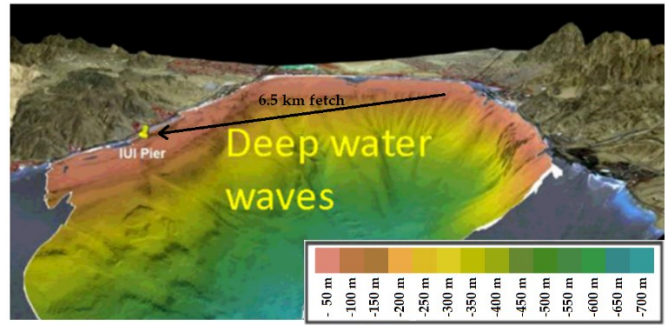

Figure 6 Topographic layout of the northern tip of the Gulf of Aqaba. The location of the pier at the Interuniversity Institute for Marine Sciences (IUI) and the 6.5 $km$ fetch along the wind flow direction are also presented. The bathymetry of the gulf is displayed in color.

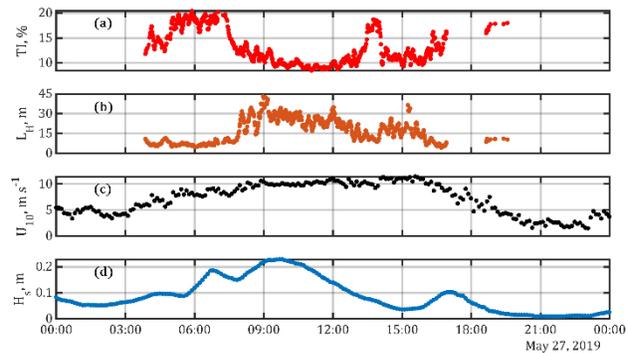

Figure 7 Temporal evolution from May 27$^{th}$, 2019 of four important wind-wave interaction parameters; (a) turbulence intensity; (b) horizontal length scale; (c) mean wind speed at ten meters above the mean water level; (d) significant wave height.

Most elements of the experimental setup in the 2019 Study are similar to those in the 2017 Study (Shani-Zerbib and Liberzon 2018; Shani-Zerbib et al. 2019). The addition to this 50-hour continuous study was the redesigned combo that was installed 4.5 m above the mean water level. The combo consisted of the same sonic, hotfilm, motor, and bridges as the laboratory experiment. Due to the harsher environmental conditions in the field, the overheat ratio was set to a lower value of 1.5. The combo provided records of the turbulent fluctuations of the wind velocity field components. The experimental setup also included an array of five Wave Staff wave gauges to sense the water wave field, two additional RM Young 81000 ultrasonic anemometers at two altitudes to obtain the mean wind speed profile, and a water pressure gauge to obtain the mean water level. A brief portion of the results is presented in Figure 7, the rest are under preparation for publication elsewhere. The temporal evolution of the four parameters presented here include the turbulence intensity, horizontal length scale, the mean wind velocity at ten meters above the mean water level, and the significant wave height. The TI and $L_H$ are defined in (1) and (3). The $U_{10}$ is the extrapolated mean wind speed at ten meters above the mean water level and the $H_s$ is the significant wave height defined as the mean wave height of the top $1/3$ of the highest waves.

The temporal evolution of these parameters reveals new patterns and raises several questions. Figure 7 reveals that the higher $L_H$ are observed only at higher $U_{10}$, also accompanied by lower $TI$ values. Additionally, the wave field appears to react to the variations in $TI$ with an increase/decrease in $L_H$ which precedes that of $H_s$. The $L_H$ are the result of the wind BL disturbance by longer waves traveling from far, approximately 6.5 $km$, and the fetch appears to limit the water-wave growth. The 2017 Study (Shani-Zerbib et al. 2018) revealed that the mean wind magnitude and direction are both important for the local water waves field evolution, but this 2019 Study reveals that also the turbulence structure is just as important. Some questions that this study raised include: is less turbulence needed to generate higher waves? What is the sensitivity of the wave field to small variations in wind flow direction? These call for more detailed experiments in the Gulf of Aqaba.

This experiment not only verified the durability and working capability of the new combo design, but also showed its capability of working continuously without human intervention in changing field conditions. This high frequency response anemometer is useful for many types of field studies in all types of environments, both changing and steady. The automated calibration approach is presented in the next section along with the previously used calibration procedures. Examination of the results will reveal the possibility of implementing or deploying the combo on moving vehicles, such as boats, cars, or heavy lifting drones for complete BL scanning.

## 2.4 Different calibration approaches

This section summarizes different calibration approaches commonly used in the literature and proposes a procedure to automate the calibration by making several changes to one of the previously used methods. Traditionally, the calibration procedure includes either a low turbulence jet or a low turbulence wind tunnel to calibrate hotwires/hotfilm. In the use of multi-wire probes, a gimbal or mechanical pitch and/or yaw manipulator is additionally necessary. The estimation of the voltage to velocity transfer function in laboratory studies is most commonly computed using the least-squares polynomial fitting using King's law or using the lookup table method (Van Dijk and Nieuwstadt 2004). However, Van Dijk (2004) (Van Dijk and Nieuwstadt 2004) argues that the most trustworthy estimation method is the lookup table, because it does not oversimplify (or underfit) the transfer function. Historically, it was not preferred as it is computationally costly, however modern computational power makes this method much more practical to implement, i.e. using the *griddedinterpolant* MATLAB® function (Tsinober et al. 1992).

For this reason, we decided to use the lookup table as the ground truth (GT) reference and to compare all attained results to the GT to quantify the performance quality of each method; a summary of all methods elaborated in this section is listed in Table 3. The GT set of the laboratory experiments is also referred to as $TJ3$ - it stands for lookup **T**able using the **J**et data with an output of **3** velocity components, i.e., no redundant information of the streamwise component to be averaged. The GT used for the field experiment is elaborated later in the section. The redundant information of the $u$ component stems from our use of a 4-wire hotfilm probe. The use of two x-probes compensates some spatial resolution but has the benefit of resolving the streamwise component at higher signal to noise ratio. Two independent studies (Van Dijk and Nieuwstadt 2004; Kit and Liberzon 2016) showed an alternative approach of averaging two sub probes that were defined a bit differently, yet neither was shown to be superior to the other. They defined them as the two x-probes, $uw_{sub1} = f(E_{1,1}, E_{1,2})$ and $uv_{sub2} = f(E_{2,1}, E_{2,2})$, and averaging only the $u$ component of the sub probes. In the text, we refer to this method as $TJ_2$; it stands for lookup **T**able using the **J**et data with an output of **2** velocity components each time, i.e., the redundant information of the streamwise component needs to be averaged.

The motivation of the selection of two x-probes was mostly economical and based on Kit and Liberzon (2016) (Kit and Liberzon 2016), where two x-probes are oriented 90 degrees from each other to provide the complete 3D flow components. Each x-probe captures two components of the velocity field, and since they are oriented at a 90° (in the roll axis) relative to each other, all three components are captured. One probe provides the $u, v$ and the other provides the $u, w$ components. The transfer function can either be computed for each x-probe separately while averaging the redundant $u$ component or to be constructed using 4 voltage inputs with 3 velocity outputs. Kit and Liberzon (2016) (Kit and Liberzon 2016) already demonstrated that neither is superior to the other and we confirm this result. Finally, as the hotfilms gradually degrade with use resulting in measurement drift, the traditional jet-based calibration was performed twice. Both the pre-measurement and post-measurement jet-based calibration sets were used to construct two independent transfer function estimates, the outputs from these two were eventually averaged to correct for the drift error.

The other traditional method commonly used for hotwire calibration is the polynomial fitting using the least-squares method. This is based on King's Law that relates the heat transfer coefficient to the fluid velocity using a polynomial approximation whose coefficients are obtained during the calibration. Van Dijk (2004) (Van Dijk and Nieuwstadt 2004) has shown that this approximation oversimplifies the heat transfer coefficient relationship to velocity, and suggested for future (and past) studies to consider only the lookup table method. When we attempted to examine this claim using our wide attacking angle calibration sets, we noticed that using 1000 different initial guesses for the least squares fit numerical algorithm provided 1000 different solutions. This most likely means that the fit converges to a local minimum with each guess, which was confirmed by calculation of the self-reconstruction errors of more than 100,000 solutions. The computational costs, however, are tremendously high making it an impractical approach for high turbulence intensity conditions characterized by a wide range of flow attack angles. In this chapter we refer to this method as $PJ3$; it stands for least-squares **P**olynomial fitting using the **J**et data with an output of **3** velocity components, i.e., no redundant information of the streamwise component to be averaged. Since the lookup table interpolates and averages between known values and does not use a single parametric curve to fit all data, we selected $TJ_3$ as GT reference.

Table 3 Examined calibration methods of the 4-wire hotfilm probe. The boxed cells correspond with the reference signals - ground truth (GT), in the laboratory and field. Each method is denoted by a two lettered acronym with a numbered index; they are all listed on the right two columns of this table. The first letter of the acronym describes the method name, i.e., **T**: lookup **T**able, **P**: **P**olynomial curve fitting, **S**: **S**hallow NN, **D**: **D**eep NN. The second letter of the acronym describes the calibration data source, i.e., **J**: **J**et obtained calibration data, **I**: **I**n-situ calibration data, **H**: **H**andpicked points from the in-situ data, **A**: **A**ll points are from the in-situ data. Finally, the numbered indices represent the number of velocity components in the calibration function output, i.e., 2 represents only two components per computation and the requirement for averaging of the redundant component, and 3 does not produce a redundant component and therefore does not require averaging.

| Calibration method | | Inputs | Outputs | Jet | In-situ |
|---|---|---|---|---|---|
| Lookup table | | $E_{1,1}\ E_{1,2}\ E_{2,1}\ E_{2,2}$ | $u, v, w$ | $GT_{lab} = TJ_3$ | $TI_3$ |
| | | $E_{1,1}\ E_{1,2}$ <br> $E_{2,1}\ E_{2,2}$ | $u, w$ <br> $u, v$ | $TJ_2$ | $TI_2$ |
| Polynomial curve fitting | | $E_{1,1}\ E_{1,2}\ E_{2,1}\ E_{2,2}$ | $u, v, w$ | $PJ_3$ | - |
| Shallow Neural Network | Handpicked training set | $E_{1,1}\ E_{1,2}\ E_{2,1}\ E_{2,2}$ | $u, v, w$ | - | $GT_{field} = SH_3$ |
| | all data for training set (automated) | $E_{1,1}\ E_{1,2}\ E_{2,1}\ E_{2,2}$ | $u, v, w$ | - | $SA_3$ |
| Deep Neural Network | all data for training set (automated) | $E_{1,1}\ E_{1,2}\ E_{2,1}\ E_{2,2}$ | $u, v, w$ | - | $DA_3$ |

The jet-based calibration methods are impractical in the field due to the above-mentioned variability of environmental conditions, and the in-situ calibration method was originally proposed as an alternative by Oncley et. al. (Oncley et al. 1996; Poulos et al. 2006). In-situ calibration was later implemented several times in the laboratory to demonstrate its capabilities (Kit and Liberzon 2016),(Kit et al. 2010) and many times in the field (Kit and Grits 2011; Fernando et al. 2015; Kit and Liberzon 2016; Kit et al. 2017; Goldshmid and Liberzon 2018). The studies in the laboratory presented the ability to use the simultaneously measured low frequency signal form the sonic to calibrate the high frequency signal of the hotfilm by means of least squares polynomial fitting and training artificial neural networks (NN). While the transfer function never sees a high frequency data, based on the understanding of the physics of heat transfer it is assumed that the heat transfer rate dependency on velocity is not a function of frequency. Hence training a transfer function model on low pass filtered data results in a model fitting to translate all frequencies similarly. The low pass filtering frequency is determined by the empirical relationship of the upper sonic trusted frequency to the mean velocity (Kaimal and Finnigan 1994),

$$f = \frac{\bar{u}}{2\pi} \ , \quad (10)$$

where $L$ is the acoustic fly-path of the ultrasonic signal depending on the sonic model in use, here $L = 0.15\ m$. The wind tunnel data collected in this study had the sonic monitoring the flow simultaneously therefore providing another calibration set inside the wind tunnel for comparison. The presentation of a polynomial curve fitting of the in-situ data would have been redundant to the studies of (Kit and Liberzon 2016),(Kit et al. 2010), it would not provide any new insights and was therefore chosen to be omitted from this study. Instead, and according to the suggestion made by Van Dijk (2004) (Van Dijk and Nieuwstadt 2004), the in-situ calibration set was used for the lookup table transfer function estimates. The two sets are $TI_3$ and $TI_2$; these stand for lookup **T**able using the **I**n-situ data with an output of **3** or **2** velocity components, respectively. Here too, the redundant information of the streamwise component doesn't or does need to be averaged, respectively. After discussing the two estimation approaches (lookup table method and the polynomial fitting) using the in-situ calibration set, the NN approach is considered.

NN have also successfully been used for complex function approximations. Several studies have shown that regression NN are capable of capturing the velocity-voltage relationship estimates strikingly well in hotwire anemometry (Kit and Grits 2011; Fernando et al. 2015; Kit and Liberzon 2016; Kit et al. 2017; Goldshmid and Liberzon 2018). In these studies, the training set (TS) used for each model was based on delicately selected minutes that contained mean velocities that covered the entire range of interest and high rms values of velocity fluctuations. Elimination of such manual selection procedure will fully automate the in-situ NN based calibration of combo data and enable the fine scale measurements to be obtained in the field independently of human decision; the automation procedure is discussed in more detail below. It mainly sums up to our suggestion of using all low pass filtered data points within a selected period of constant ambient properties, usually an hour, as the TS. As in the previous methodology for the in-situ calibration of the combo probe, the use of multilayer feedforward NN is implemented using the MATLAB Deep Learning Toolkit.

The motivation to use deep learning stems from the fact that advances in machine learning research teach us that in order to achieve better training of NN, the TS should be as representative as possible to describe the whole dataset (Ng 2018). We, therefore, propose to use all low pass filtered data of a predetermined period with constant ambient properties as the TS. While inclusion of all low pass filtered data in the TS best represents the dataset, it increases the TS size by at least an order of magnitude and requires an increase of the network size to avoid underfitting. Moreover, as the selection of an hour was previously chosen arbitrarily because of the flows examined, the general requirement is for the background flow conditions to remain constant, i.e., temperature, pressure, and humidity. The automated process should include a maximum duration and a range in which these properties are considered sufficiently constant. The low pass filtering frequency should be chosen independently for each minute based on the mean velocity of that minute using the empirical relationship (10). To quantify the automated procedure results (item 3 in the list below), we also computed the previously implemented (Kit et al. 2010, 2017; Kit and Grits 2011; Vitkin et al. 2014; Fernando et al. 2015; Kit and Liberzon 2016; Goldshmid and Liberzon 2018, 2020) NN procedure (item 1 in the list below) and added an intermediate step as well (item 2 in the list below). A list of the NN based methods examined includes:

1. The $SH_3$; this stands for **S**hallow NN (2 layers) with 100 hidden neurons, a **H**andpicked TS, and a **3**-component output—therefore no redundant components for averaging (this is the same for all three NN models). The selection of the TS was conducted separately for each flow type, allowing a separate model to be trained. This is also the GT for the field experiment.
2. The second shallow NN method we examine is the $SA_3$; this stands for a **S**hallow NN (2 layers) with 100 hidden neurons, but with an automated TS using **A**ll data points collected with similar ambient conditions. This is an extreme case because three different flow types attained in the laboratory (listed in Table 1) are examined together and calibration transfer functions are constructed using the same model.
3. Finally, the third NN model type we examine is $DA_3$; this stands for **D**eep NN model (15 layers) with **A**ll low pass filtered data included. Each layer has 100 hidden neurons, and the TS is also selected using all data points from all three flow types. This is also the *suggested method* for complete automation.

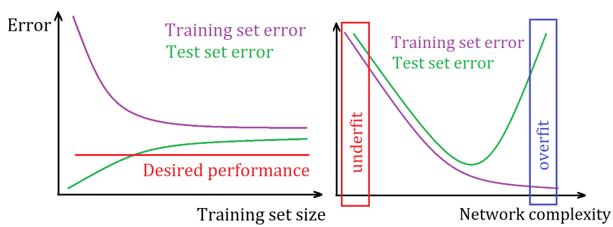

Figure 8 Left: Example error diagnosis of training a NN model with respect to TS size (Ng 2018). An increase in the size of the TS will not necessarily provide the desired performance. Right: Error diagnosis example of bias vs variance in training a NN model. The underfitting or overfitting (bias vs variance) of a model can be modified based on the network complexity.

To eliminate avoidable errors, we trained 10 separate NN models with similar hyperparameters and average their results; this is true for all NN results we discuss in this study. The only difference between the 10 NN models is the random split of the training/testing data points from the TS. We noticed that the large scales velocity fluctuations were captured somewhat differently in each network. This was observed before (Kit and Liberzon 2016), and a correction was suggested using the large scales from the sonic. This difference is exhibited as a constant shift of the power density spectrum of velocity fluctuations, although the shape remained the same. When averaging the 10 NN together, it was noticed that the deviation of the large scales from those provided by the sonic became smaller. It was suggested that the final averaged time-series should be shifted to correspond with the trusted large scales of the sonic. The inconsistency between the 10 NN outputs raised a question of whether such bias is avoidable. After noticing that the polynomial fits using various initial guesses converged to different solutions (probably reaching local minima), we suspected that the NN might be doing the same thing. An attempt to tackle this was made using the training of a deep NN. It is suggested (Kawaguchi 2016) that deep NN are highly unlikely to have local minima, instead saddle points are expected and the global minimum is more susceptible to be found when a solution is reached. Indeed, after training the deep NN the scatter in large scales results between the individual NN realization was avoided. This result bolsters the above assumption that the shallow NN was too simple to represent the complex voltage to velocity transfer function.

Selection of 15 layers for the deep network was dictated by the memory limits of the used PC. Our initial suggestion for automation of the in-situ NN based training is in increasing the TS size by using all low pass filtered data. The motivation is to maximize the variability of voltage-to-velocity conversion data to train the network. Figure 8 shows (Ng 2018) that an increase in the TS size might not be sufficient to obtain the desired performance and an increase in the complexity of the network should be considered as well. This is also known as the bias vs variance tradeoff in machine learning. The two most common sources of possible errors in NN training are bias and variance (Ng 2018). The variance (associated with overfitting) of a trained NN is how much worse the model performs on unseen data than it did on the TS. If the variance is high, it is commonly helpful to increase the TS size. Bias (associated with underfitting) is the embedded error in the TS and is broken down to avoidable and unavoidable bias. An example of unavoidable bias is the sonic measurement limitation, while an example of avoidable bias is a simplified model to represent a complex relationship, Kings law in our case. The avoidable bias can be eliminated by increasing the size of the model (i.e., adding more neurons and layers). Theoretically, when the bias is small, but the variance is large, adding more training data will probably help close the gap between the test and TS errors. Eventually we increased both the TS size and the model complexity, and the difference between $SA_3$ and $DA_3$ will present the effects of each of the steps.

We did not perform a complete learning curve analysis on this set because that might not be useful to implement across different flows when attempting to automate the calibration procedure of the combo probe in field studies. The other reason is that we do not have a real ground truth for the reference, using the sonic data as a target our TS has an embedded unavoidable error in it. The sonic sensed velocity field are of finite accuracy, set by the manufacturer calibration correcting for the presence of struts and the use of sophisticated electronics to resolve the acoustic signal Doppler effect measurements. Moreover, the sonic provides spatially averaged measurements, depending on their acoustic signal fly-path length, in the case of the used here wind tunnel measurements the ratio of the sonic acoustic fly path to the BL height is about 20%. While in the field it would be at least 1-2 orders of magnitude smaller and is expected to perform much better.

To conclude, this section discussed various calibration methods, some of which are commonly used, and a newly proposed one. The performance of these methods in terms of small scales, the large scales, and turbulence statistics comparisons will be examined in the next section.

## 3 Results and Discussion

This section discusses the performance of the methods for estimating the voltage to velocity transfer function. These methods include both traditional approaches and the newly proposed automated approach. A comparison between the signals reconstructed using different calibration approached is made. More specifically, quantification of a normalized errors in reconstructing both the small and large scales are presented along with a qualitative comparison of the spectral shapes. These present the validity of using the automated method and its robustness and capabilities.

The qualitative comparison of the small scales is achieved using the average delta error. It is defined as the average over errors of all three velocity components,

$$\bar{\delta} = \frac{(\delta_u + \delta_v + \delta_w)}{3}, \qquad (11)$$

similarly to how it was defined (Vitkin et al. 2014; Kit and Liberzon 2016). The delta error is defined as the normalized to the rms, and therefore the non-dimensional error parameter (Vitkin et al. 2014; Kit and Liberzon 2016), defined for each velocity component separately, is

$$\delta_u = rms\left(\frac{u}{rms(u)} - \frac{u_{GT}}{rms(u_{GT})}\right), \qquad (12)$$

$$\delta_v = rms\left(\frac{v}{rms(v)} - \frac{v_{GT}}{rms(v_{GT})}\right), \qquad (13)$$

$$\delta_w = rms\left(\frac{w}{rms(w)} - \frac{w_{GT}}{rms(w_{GT})}\right). \qquad (14)$$

The $u, v, w$ represent the fluctuations of the signal in the streamwise, longitudinal, and transverse components of velocity. The subscript $GT$ represents the ground truth or reference signal, i.e., $TJ_3$ in the laboratory and $SH_3$ in the field. These $\delta$ errors are presented in Figure 9, separately for each flow regime and as a function of the mean velocity. In the field, the $\delta$ values of the proposed automated method ($DA_3$) are presented separately for each velocity component. All four flow regimes present a general reduction of error with an increase in the mean velocity. They are also of the same range observed in the previous study (Kit and Liberzon 2016). This is expected for all sonic based calibration procedures, as the sonic performance in resolving the large scales improves with an increase in mean velocity (Kaimal and Finnigan 1994). In the laboratory measurements, the jet-based calibrations seem to depend less on the mean wind speed. The $\bar{\delta}$ values of $TJ_2$ are the smallest because the comparison of all methods is against the $GT$

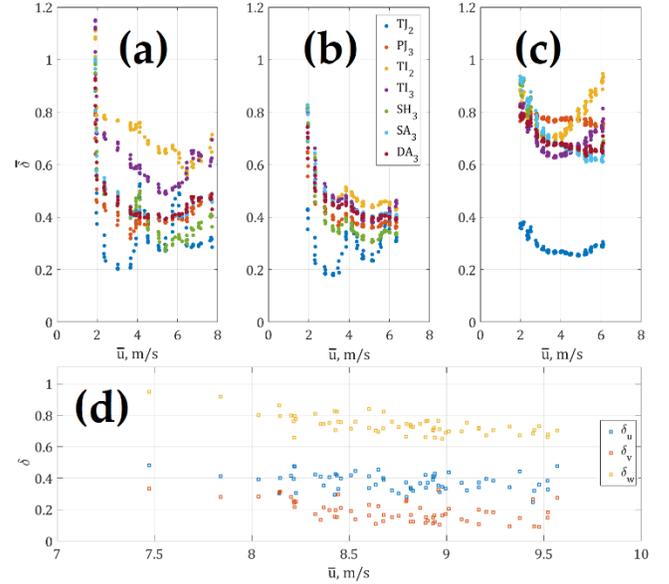

Figure 9 (a-c) Average delta error, $\bar{\delta}$, for all calibration procedures examined here as a function of the mean streamwise velocity. All delta errors are with respect to the previously selected ground truth $TJ_3$. These are broken down by the flow regime examined (a) NGNC, (b) YGNC, and (c) YGYC. (d) The delta error for all three velocity components with respect to the mean streamwise velocity in a representative hour at 9 am on May 27$^{th}$, 2019.

which is the $TJ_3$ set in this case, and these are expected to provide the most similar transfer function estimates. The key takeaway from the $\delta$ analysis is that the range of $\delta$ is the same for all methods. The small scales reconstruction errors are expected to translate to errors in the calculated quantities such as length scales, dissipation rates, and turbulence intensities. Hence, some variability of the latter between the calibration methods is expected. This was observed in all the flow regimes examined (Goldshmid 2020).

Next, is a qualitative observation of the spectral shapes of each flow regime examined in the lab; Figure 10 presents the average spectral shapes of all minutes at the highest wind speed for each flow regime, separately for each velocity component. The spectra were computed averaging over one second windows, resulting in 1 $Hz$ spectral resolution, and each curve represents a different calibration method. In the ideal case, all spectral curves in each flow regime should be identical, regardless of the calibration method. The clearly visible "bump" in the spectral shapes of the YGYC flow regime is attributed to the presence of the canopy; it is not seen in any of the other examined flows here. The blue $-5/3$ line is placed at the same location in all figures to easily distinguish between the types of flows. The fact that the there is a significant change in the average intensities between the flows indicates that the hotwires can differentiate between the different flows examined using *all* examined calibration procedures. In practice, this is the key/critical point that is sought for in the experiments. Such ability to distinguish between drastically changing flow regimes, even if the actual values are recorded with some error, is crucial especially for field implementation where high variability of the conditions is expected and unavoidable.

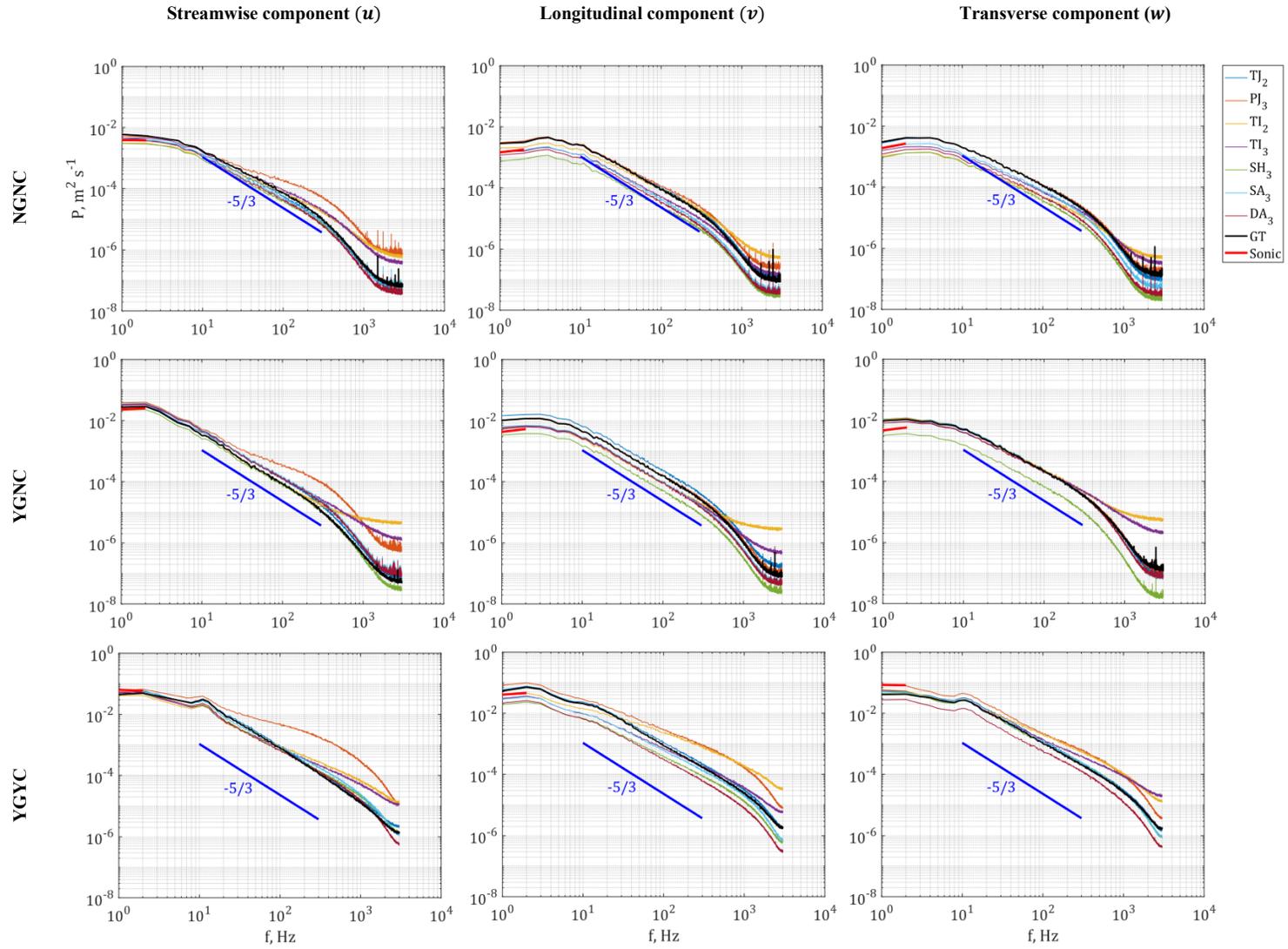

Figure 10 Averaged power density spectral curves of the streamwise (row 1), longitudinal (row 2), and transverse (row 3) velocity components fluctuations of all data obtained at the highest flow rate of the NGNC (column 1), YGNC (column 2), YGYC (column 3) flow regimes. All blue -5/3 lines are placed at identical location on all plots. The red lines are spectral data from the sonic, up to the trusted frequency limit.

The differences between the spectral shapes are more pronounced in the smaller scales than in the larger scales. For the laboratory data, the small scales of the in-situ calibrated lookup table method ($TI_2, TI_3$) and the jet calibrated polynomial ($PJ_3$) seem to suffer from excessive noise, consistently appearing more elevated than in the $GT$ and the NN counterparts. Both the $GT, SA_3$ and $DA_3$ exhibit a similar shape in in all three flow regimes and velocity components, with one exception: the $u$ components in the YGYC flow. The $DA_3$ might have been able to capture the dissipation range in the $2-3\ kHz$, where the $GT$ and $SA_3$ methods seem to have captured noise as their curves flatten horizontally at the high frequencies. As mentioned in the previous section, each NN model consists of 10 separate models whose final outputs are averaged. It was also observed that the $SA_3$ had a larger range of variations between the outputs of the 10 models, and $DA_3$ had a much smaller range. Based on these findings we can recommend the $DA_3$ approach for the calibration automation of the combo probe.

Ideally, a well-trained NN model should only capture the coherence of a signal and miss/omit any noise. Therefore, there is the possibility that the hotwires are solving the larger scales more precisely than the sonic in the laboratory experiments. This is especially true because of the relatively large acoustic fly path of the sonic relative to the height of the BL in the lab. Since we do not have any additional reference to compare with, such as PIV, no concrete conclusions can be drawn here, and this issue should be examined further in future studies. Our approach is to quantify the large-scale deviation relative to those of the $GT$.

Originally, we selected the most "trusted" method as the standard calibration method using the automatic calibrator and deducing the transfer function using the lookup table ($TJ_3$). Here, we will compare the absolute difference of the sonic provided large scales from those of the selected ground truth. Then we will compare how those differences relate to the outputs provided by the three NN based calibration methodologies. This is to test whether the hotfilm can capture the large scales more accurately than the sonic. If so, the correction that was previously suggested (Kit and Liberzon 2016) might no longer be necessary. These normalized differences are:

$$\Delta P_{NN} = \frac{|P_{NN} - P_{GT}|}{P_{GT}}, \quad (15)$$

and,

$$\Delta P_{Sonic} = \frac{|P_{Sonic} - P_{GT}|}{P_{GT}}. \quad (16)$$

For consistency, the comparison is made up to $2Hz$ since it is the highest trusted frequency of the sonic at the lowest mean velocity examined in this study. The comparison is only made with the NN based methods because these were previously recommended to undergo a correction based on the sonic reading of the large scales (Kit and Liberzon 2016). Figure 11 details the results per flow regime.

In the case the above discussed correction would be unarguably necessary, the cloud of points should have been densely stacked in the southwest corner of all figures. Since this is not the case, the possibility that the hotwires resolve the large scales more accurately than the sonic should be considered. All methods fall close to the $1:1$ ratio line. We are, essentially, comparing the large scales of the sonic to the large scales of the NN derived signals that are based on the sonic data. The comparison and normalization of both are based on the $GT_{lab}$ large scales; these signals are calibrated using the jet and are unrelated to the sonic signals. Since there are not many points that lie below the $1:1$ line, it appears that the uncorrected NN large scale signals using the sonic data for calibration ultimately have smaller differences from the $GT_{lab}$ than the sonic raw large-scale signals. Indicating the NN bases calibration is more accurate than the sonic in reconstructing the large-scale fluctuations.

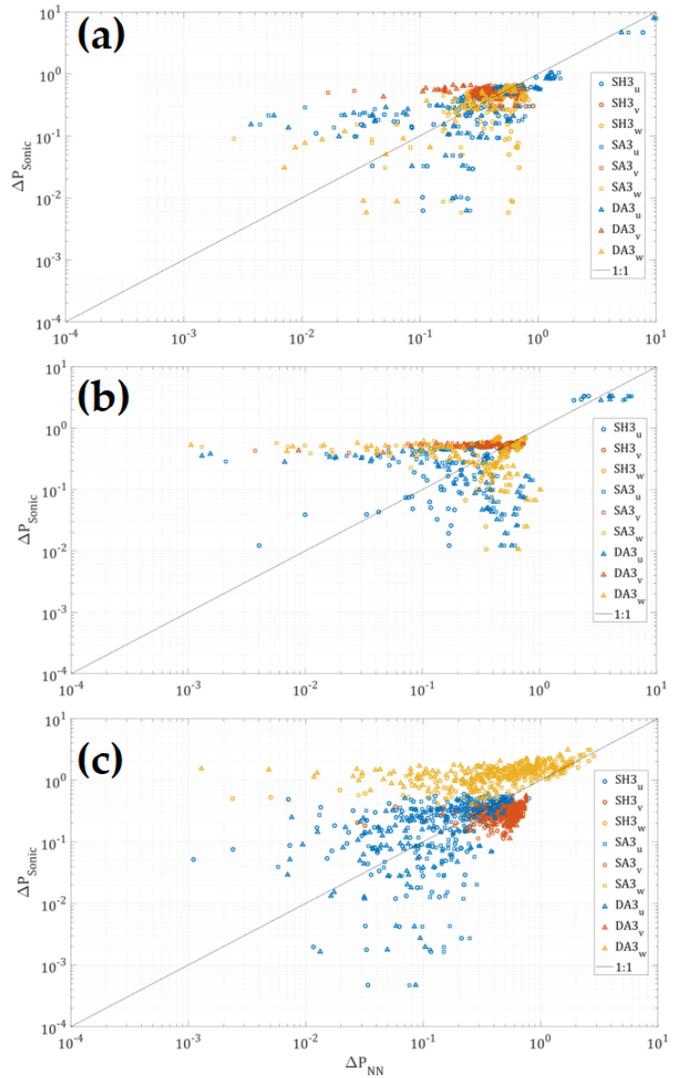

Figure 11 The **(a)** NGNC, **(b)** YGNC, **(c)** YGYC flow regimes large scale normalized deviation of the NN methods results relative to those of the sonic. All compared against the selected GT dataset. The circles, squares and triangles represent the $SH_3, SA_3, DA_3$ sets correspondingly. The blue, orange, and yellow represent the $u, v, w$ components correspondingly.

The here presented tests were conducted under a range of acceptable uncertainty. The transfer function estimates are expected to vary up to a certain degree depending on the implemented calibration and the sensors averaging methods were. For example, the lookup table method using two or four sub probes exhibits some difference in their results. Although out of the scope of this work and planned to be addressed later, the jet calibration data used with the deep NN also demonstrated great performance and can be implemented as a method for estimating the voltage to velocity transfer functions in laboratory studies because of its robustness.

In conclusion, the large to small scale performance of each calibration method was examined using both laboratory and field obtained data. All methods presented $\delta$ value of the same range, supporting the validity of the proposed automated NN based calibration procedure - $DA_3$. The most complex/extreme case examined in the lab included combining all three different flow regimes into one large dataset for the training of a single transfer function. In the actual field measurement, when the conditions change this much, a new model would be trained. This goes to show the robustness of the deep NN model and its capability to find the coherence even in highly variable flow conditions. The automation of the calibration procedure enables near real-time monitoring of fine-scale turbulent fluctuations in the field.

## 4 Conclusions

Prior to the combo, hotwire anemometry was rarely used in the field because of the need for cumbersome and frequent re-calibration and for constant re-alignment with the mean wind direction due to variability of environmental conditions. The combo was previously shown to tackle these limitations (Fernando et al. 2015; Kit et al. 2017; Goldshmid and Liberzon 2018, 2020) by combining the hotfilm probes with a low spatiotemporal resolution instrument – sonic, and implementing an NN based in-situ calibration method. It is capable of continuously sensing the fine scales of turbulent fluctuations in the field with high accuracy. This study presented the most recent improvements made to the combo mechanical design and data processing procedures. The main findings of this study can be summarized as follows:

1. The new mechanical design removes the restriction of the combo measurements at 120° range. Rigidly connecting the hotfilms sensors to the sonic and rotating both now enables measurements of the flow at any angle of attack in a full 360° range.
2. To examine the durability of this design, the combo was deployed in an open sea environment for several days. This experiment confirmed that the combo operates properly and has the capability of operating continuously without human intervention for days or weeks at a time. Although the data collection prior to this study did not require any human intervention, the calibration procedure itself was human decision based.
3. The automation of the combo calibration procedure removed potential human errors and enabled near real-time monitoring of fine-scale turbulent fluctuations. The proposed automation procedure was tested on both the open sea dataset and inside a wind tunnel. A comparison of the new procedure results was made against traditional calibration methods using a low turbulence intensity jet and a mechanical manipulator. The automated procedure included the use of a deep NN, instead of the previously used shallow NN, and the use of all low pass filtered data of steady ambient conditions for the training set. The automated calibration in the wind tunnel was used to represent an extreme case of significant variations in wind magnitude and turbulence characteristics during the span of a measurement. Here the sonic provided large scale velocities fluctuations data of three different turbulent BL flow regimes were used together to train one transfer function for all flow regimes. The discussed results confirmed that the new NN training set approach allows combo to successfully produce measurements over even extreme changes in the flow conditions.
4. The deep NN appeared to have much higher frequency response and hence capture better the finest scales, as is evidenced in the examined spectral shapes of velocity field components fluctuations power density. They exhibited a continued dissipation rage at much higher frequencies instead of the flattening observed by other methods. As for the field data, the results exhibited the same range of differences between the automated procedure using deep NN and the human-decision based calibration method using shallow NN.

The automated training was introduced in an attempt is to minimize the avoidable human based bias. Our results do not allow a conclusion regarding the ultimately most accurate method, but we can say with certainty that the accuracies achieved by the commonly used and the newly proposed methods are comparable. Meaning the combo *can* and *should* be implemented in the field and the automated calibration procedure using deep NN is preferable.

The combined results of this study show that accurate measurements of atmospheric turbulent flows in the field are now more achievable than ever before. The new, and mechanically simpler, combo design makes this possible in all 360°. The goal of any turbulent flow monitoring is to identify relative changes in turbulent properties, regardless of the selected procedure for estimating the voltage-to-velocity transfer function. The combo with the automated calibration procedure has shown to manage this task appropriately. This high spatiotemporal resolution anemometer is useful for many types of field studies across wide range of environmental conditions, both steady and highly variable in terms of mean flow parameters and turbulence characteristics. The automated procedure enables almost real-time processing, which would provide stationary meteorological stations the ability to monitor real-time fine-scale turbulence statistics. This is especially useful with changing field conditions and consequently with non-stationary measuring stations, such as probes placed on moving platforms such as moving vehicles, boats, and drones that can be used to scan the entire BL.

**Author Contributions:** R.G. and D.L. conceptualized the research; R.G., D.L., and E.W. designed the experiments; R.G and E.W. performed the experiments; R.G. analyzed the data; R.G. and D.L. contributed to data collection and analysis tools and software; D.L. was responsible for funding acquisition; writing-review and editing was performed by R.G and D.L..

**Acknowledgements**. We express our warmest gratitude to Ran Soffer for helping us with the wind tunnel experiments and Eva Chetrit for helping with the assembly of the new combo design. The water waves portion of the experimental setup discussed in Section 3.2 was conducted in collaboration with Almog Shani-Zerbib. The authors acknowledge the support of the Israel Science Foundation grant Nº 2063/19.